\documentclass[11pt,a4paper]{article}
\usepackage{amsmath,amssymb,mathrsfs}
\usepackage{graphicx}
\usepackage{cancel, comment}

\newcommand{\vev}[1]{\left\langle #1\right\rangle}

\newcommand{\ms}[1]{\mathscr{#1}}
\newcommand{\mc}[1]{\mathcal{#1}}

\begin{document}
\begin{titlepage}
\begin{center}

February 2015 \hfill IPMU14-0293\\
\hfill UT-14-36

\noindent
\vskip3.0cm
{\LARGE \bf 

Generic Scalar Potentials for Inflation in \\ Supergravity with a Single Chiral Superfield

}

\vglue.3in

{\large
Sergei V. Ketov~${}^{a,b,c}$ and Takahiro Terada~${}^{d}$ 
}

\vglue.1in

{\em
${}^a$~Department of Physics, Tokyo Metropolitan University \\
1-1 Minami-ohsawa, Hachioji, Tokyo 192-0397, Japan \\
${}^b$~Kavli Institute for the Physics and Mathematics of the Universe (IPMU)
\\The University of Tokyo, 5-1-5 Kashiwanoha, Kashiwa, Chiba 277-8583, Japan \\
${}^c$~Institute of Physics and Technology, Tomsk Polytechnic University\\
30 Lenin Ave., Tomsk 634050, Russian Federation \\
${}^d$~Department of Physics, The University of Tokyo,\\
7-3-1 Hongo, Bunkyo, Tokyo 113-0033, Japan
}

\vglue.1in
ketov@tmu.ac.jp, takahiro@hep-th.phys.s.u-tokyo.ac.jp

\end{center}

\vglue.3in

\begin{abstract}
\noindent We propose a large class of supergravity models in terms of  a single chiral matter 
superfield, leading to (almost) arbitrary single-field inflationary scalar potentials similar to 
the $F$-term in rigid supersymmetry. Those scalar potentials are positively semi-definite (in our approximation), and can preserve supersymmetry at the end of inflation.  The only scalar superpartner of inflaton is stabilized by a higher-dimensional term in the K\"{a}hler potential. We argue that couplings of the inflaton to other sectors of the particle spectrum do not  affect the inflationary dynamics, and briefly discuss reheating of the universe
by the inflaton decays. 
\end{abstract}
\end{titlepage}

\section{Introduction}\label{sec:intro}

Inflationary cosmology~\cite{Starobinsky:1980te, old_inflation, new_inflation} is now getting established by the recent precise observations of the Universe such as the WMAP~\cite{Hinshaw:2012aka} and the Planck~\cite{Ade:2013uln}.
For example, the spectral index and the tensor-to-scalar ratio are constrained by the Planck+WP+highL+BAO to $n_{\text{s}}=0.9608\pm 0.0054$ at 68\% CL and  $r<0.111$ at $95\%$ CL, respectively. 
On the other hand, the BICEP2 claimed that they discovered $r=0.16^{+0.06}_{-0.05}$ after foreground subtraction with $r=0$ disfavored at $5.9\sigma$~\cite{Ade:2014xna}.
The large value of $r$ claimed by the BICEP2 implies a large-field inflation because of the Lyth bound~\cite{Lyth:1996im}.  By the way, the renowned Starobinsky model~\cite{Starobinsky:1980te} is fully consistent with the Planck data by
predicting a very small tensor-to-scalar ratio $r\simeq 4\times 10^{-3}$, being a large-field inflationary model also. There are some arguments in the literature~\cite{Mortonson:2014bja, Flauger:2014qra, Adam:2014bub} that the BICEP2
collaboration underestimated the foreground, so that $r\simeq 0$ may still be consistent with the data.
Therefore, the inflationary models with $r\lesssim 0.1$ are still alive in the present situation. The actual value of
$r$ is going to be established by new observations in a not so distant future.

Under such circumstances, we are interested in embedding the inflationary models consistent with current observations into a more general framework motivated by particle physics and a fundamental theory of quantum gravity such as superstrings or M-theory. It is natural to consider supergravity~\cite{Freedman:1976xh, Deser:1976eh, WessBagger} for that purpose because (i) supergravity emerges as the low-energy effective action of
superstrings, and (ii) the energy scale of inflation is higher than the electroweak scale but is lower than the Planck scale where some unknown UV effects may come in. We pursue {\it minimal} realizations of inflation in supergravity, by minimizing a number of the matter d.o.f. involved.

Describing inflation and, in particular, a large field or chaotic inflation~\cite{Linde:1983gd} in supergravity is known to be {\it non-trivial}, because of the presence of the exponential factor and the negatively definite term in the $F$-type scalar potential. A shift symmetry of the K\"ahler
potential plays the crucial role in the model building of chaotic inflation in supergravity~\cite{Kawasaki:2000yn}. In those pioneering papers yet another chiral superfield (sometimes called the (s)Goldstino or Polonyi superfield) of the $R$-charge $2$, with the vanishing  vacuum expectation value, was introduced to allow a positively definite inflationary scalar potential. Later on, some extensions of that idea with more general chaotic inflationary potentials were introduced in supergravity in Refs.~\cite{Kallosh:2010ug, Kallosh:2010xz, Kallosh:2011qk}.  
A different approach for inflation in supergravity with vector or tensor supermultiplets was proposed in Ref.~\cite{Farakos:2013cqa}, and it was extended to embed arbitrary scalar potentials in  Refs.~\cite{Ferrara:2013rsa, Ferrara:2013kca}.  All those methods employ the second superfield, in addition to that containing inflaton. 

Inflation with a single superfield, or ``sGoldstino inflation'', was previously studied in Refs.~\cite{AlvarezGaume:2010rt, Achucarro:2012hg}, and it was concluded that a large-field inflation is virtually impossible
in that case \cite{Achucarro:2012hg} (see also~\cite{Roest:2013aoa}). Recently, new models with a nilpotent chiral 
 Goldstino superfield were proposed~\cite{Antoniadis:2014oya, Ferrara:2014kva}, which lead to the standard Volkov-Akulov action for Goldstino in nonlinearly realized supersymmetry, and a large field inflation is possible.
Though those models have only one dynamical complex scalar, their fermionic sectors are much more complicated.
In this paper we adopt the standard (linearly realized) supergravity, with only one chiral (inflaton) superfield, eluding the
known no-go statements. It is worth mentioning here that there is another minimal scenario in which inflation is driven by gravitino condensation~\cite{Ellis:2013zsa}, though with the use of a dilaton chiral superfield in conformal supergravity,
in order to make gravitino lighter than the Planck scale.

In our previous short paper~\cite{Ketov:2014qha} we proposed some new supergravity models, realizing a large field 
inflation and extending the quadratic~\cite{Linde:1983gd} and the 
Starobinsky~\cite{Starobinsky:1980te} models, by using a single (inflaton) chiral superfield only, in the standard supergravity.
The required degrees of freedom, other than those of the standard supergravity including graviton and gravitino, were reduced by half from those available in the literature where either an extra chiral~\cite{Kawasaki:2000yn, Kallosh:2010ug, Kallosh:2010xz, Kallosh:2011qk} or an extra vector~\cite{Farakos:2013cqa, Ferrara:2013rsa, Ferrara:2013kca} superfield are required, in addition to the inflaton supermultiplet.
A discovery of the fact that a large field inflation is possible in supergravity with a {\it single} chiral superfield was exciting, though the scalar potential, which we obtained in a very straightforward way, was not very transparent or illuminating enough, in contrast to the simple scalar potentials of Refs.~\cite{Kallosh:2010ug, Kallosh:2010xz, Kallosh:2011qk} and Refs.~\cite{Ferrara:2013rsa, Ferrara:2013kca}.
The reason is that the suitable K\"{a}hler potentials and superpotentials were found by a trial and error procedure in Ref.~\cite{Ketov:2014qha}. Moreover, {\it supersymmetry} (SUSY) was broken in the vacuum at the inflationary scale in some of the models presented in Ref.~\cite{Ketov:2014qha}. Though it is not necessarily a problem, it may be inconsistent with the low-energy SUSY scenario incorporating the gauge coupling unification and reducing the hierarchy problem of scalar masses.

The purpose of this paper is to present a new, special and much larger class of the minimal models, employing a single chiral superfield, which (i) lead to  very simple and (almost) arbitrary scalar potentials, 
and (ii) preserve SUSY at the end of inflation. We show that it is possible to get a single field scalar potential, like the one of global SUSY $F$-term, in the proposed class of supergravity models when using the stabilization mechanism to be explained in Sec.~\ref{sec:stabilization}.

The paper is organized as follows. In the next (and main) Sec.~\ref{sec:ArbPot} we propose a class of models that lead to nearly arbitrary positively semi-definite inflationary potentials, by using only a single chiral superfield. We give further support to our framework by arguing about viability of the stabilization mechanism and stability of inflation against possible inflaton couplings to other sectors of the theory in Secs.~\ref{sec:stabilization} and  \ref{sec:coupling}, respectively. In
 Sec.~\ref{sec:conclusion} we summarize our results. More studies of the viability of our stabilization mechanism for various K\"{a}hler potentials are given in 
 Appendix~\ref{sec:VariousK}. Some variations of the Starobinsky scalar potential in supergravity with a single chiral superfield are collected in Appendix~\ref{sec:defStarobinsky}. In Appendix \ref{sec:origin}
  we propose a mechanism 
 for the possible origin of the key quartic term in the K\"ahler potential via integration of heavy superfields.

\section{Designing arbitrary inflationary potentials in supergravity}\label{sec:ArbPot}

In Ref.~\cite{Ketov:2014qha} our method and results were not practical enough, in order to derive explicit scalar potentials suitable for inflation, because the scalar potential derived from our K\"{a}hler- and super- potentials of supergravity had a complicated form.

One of the technical reasons was the real part $\text{Re}\Phi$ that was effectively fixed to a non-zero value $\Phi_0$.
It is actually more convenient to redefine the superfield so that the vacuum expectation value (VEV) of its leading scalar field component vanishes. Let us  consider the following K\"{a}hler potential:~\footnote{ 
We take the units where the reduced Planck mass is set to one $(M_{\rm G}=M_{\text{P}}/\sqrt{8\pi}=1)$ unless it is otherwise stated.}
\begin{align}
K=&-3 \ln \left[ 1+ \left(  \Phi + \bar{\Phi}  \right)/ \sqrt{3} \right]. \label{K1}
\end{align}
It has a shift symmetry, $\Phi \rightarrow \Phi + i a$, with a real parameter $a$.
We implicitly assume here that there is a stabilization term, like $\zeta \left( \Phi+\bar{\Phi} \right)^4$, under the logarithm (it is discussed at length in Sec.~\ref{sec:stabilization}). The square root factor is introduced to obtain the canonically normalized kinetic terms. The kinetic term and the scalar potential are
\begin{align}
\mc{L}_{\text{kin}}=& -  \left( 1+\left( \Phi +\bar{\Phi} \right)/\sqrt{3} \right)^{-2} \partial_{\mu}\bar{\Phi}\partial^{\mu}\Phi, \\
V=& \left( 1+\left( \Phi +\bar{\Phi} \right)/\sqrt{3} \right)^{-3} \left( \left( 1+\left( \Phi +\bar{\Phi} \right)/\sqrt{3} \right)^{2} \left| W_{\Phi} - \frac{\sqrt{3}}{ 1+\left( \Phi +\bar{\Phi} \right)/\sqrt{3}} W \right|^2 -3 \left| W \right|^2  \right)~,
\end{align}
respectively. When setting $\vev{\text{Re}\Phi}=0$, they are simplified to 
\begin{align}
\mc{L}_{\text{kin}}=& -  \partial_{\mu}\bar{\Phi}\partial^{\mu}\Phi, \\
V=& \left| W_{\Phi} \right|^2 -\sqrt{3 } \left( \bar{W}W_{\Phi}+W\bar{W}_{\bar{\Phi}}  \right) . \label{potential0}
\end{align}

Let us now consider the superpotentials having the form
\begin{align}
W(\Phi)=\frac{1}{\sqrt{2}}\tilde{W}(-\sqrt{2}i \Phi),
\end{align} 
where $\tilde{W}$ is a real function of its argument~\cite{Kallosh:2010ug, Kallosh:2010xz, Kallosh:2011qk} (i.e. all coefficients of its Taylor expansion are real) up to an overall phase that is unphysical. 
Then the scalar potential is greatly simplified to
\begin{align}
V=|W_{\Phi}(i \text{Im}\Phi)|^2 = \left( \tilde{W}'(\chi) \right)^2 , \label{potential}
\end{align}
where the prime denotes differentiation with respect to its argument, and $\chi=\sqrt{2} \text{Im}\Phi$ is the canonically normalized inflaton field.
Demanding a real function $\tilde{W}$ may look like a strong condition, but it is obviously satisfied in the case of a monomial superpotential that is sufficient for the simplest chaotic model with a quadratic potential.
On the one hand, even if the reality or phase alignment condition is not the case, inflation could occur with a straightforwardly obtained but a little more complicated scalar potential~\eqref{potential0}.  
On the other hand, if it is satisfied at a high-energy scale, the functional form of $\tilde{W}$ is preserved by the non-renormalization theorem~\cite{non-renormalization}.

The superpotential can be viewed as a small explicit breaking of the shift symmetry in the sense of 't Hooft~\cite{'tHooft:1979bh}. Hence, one expects the shift symmetry breaking terms to be suppressed by the same scale as the superpotential also in the K\"{a}hler potential due to quantum corrections.  Those effects are beyond the scope of this paper since we are interested in a simple classical framework in the first place. See, however, Refs.~\cite{Li:2013nfa, Harigaya:2014qza} for studies of those 
extra contributions.

It is also possible to switch the roles of the real and imaginary parts by a field redefinition. When we start with
\begin{align}
K=& -3 \ln \left( 1+ \left( -i \Phi+ i \bar{\Phi} \right)/\sqrt{3} \right) \label{K2}
\end{align}
and a superpotential $W$ with real coefficients, after stabilization of the imaginary part $\vev{\text{Im}\Phi}=0$, the scalar potential reads
\begin{align}
V=&|W_{\Phi}(\text{Re}\Phi)|^2= \left( \tilde{W}' \left( \phi  \right) \right)^2, \label{potential2}
\end{align}
where  $\phi=\sqrt{2}\text{Re}\Phi$ is the canonically normalized inflaton field, and the real function $\tilde{W}$ is defined by $W(\Phi )=\frac{1}{\sqrt{2}}\tilde{W}(\sqrt{2}\Phi)$.

This way a very large class of inflationary potentials can be obtained by using only one chiral superfield.  The only restriction is that the scalar potential should be square of some real function.
And it is automatically positively semi-definite that is quite comfortable for phenomenological purposes. It is not difficult to obtain a vacuum with the vanishing cosmological constant also.
After that it is always possible to add a constant to the superpotential in order to cancel a SUSY breaking. Adding a constant to the superpotential does not affect the scalar potential because the latter is determined by the derivative of the former.

Our approach to the inflationary model building in supergravity is as powerful as those of Refs.~\cite{Kallosh:2010ug, Kallosh:2010xz, Kallosh:2011qk, Ferrara:2013rsa, Ferrara:2013kca} in the sense that the superpotential leading to an arbitrary positively semi-definite scalar potential can be approximately reconstructed by taking its square root, Taylor expanding it, and then integrating.
At the same time, our method is more economical in the sense that only a single chiral superfield (other than the standard gravitational multiplet containing graviton and gravitino) is used.

\paragraph{Example 1: a monomial potential.}
The $2n$-th power monomial potential $V=|c_{2n}|^2 \phi^{2n}$ follows from the following superpotential (in the notation of Eqs.~\eqref{K2} and \eqref{potential2}):
\begin{align}
W=\frac{2^{n/2}c_{2n}}{n+1} \Phi^{n+1}.
\end{align}
In particular, a quadratic potential $V=m^2\phi^2/2$ is obtained from the superpotential
\begin{align}
W=\frac{1}{2}m \Phi^2.
\end{align}

\paragraph{Example 2: the Starobinsky potential.}
The Starobinsky inflationary scalar potential is reproduced by the superpotential
\begin{align}
W=\frac{\sqrt{3}}{2} m \left( \Phi + \frac{\sqrt{3}}{2} \left( e^{- 2\Phi / \sqrt{3}} -1 \right) \right). \label{W_Starobinsky}
\end{align}
When using the current framework, it's easy to obtain a set of the deformed Starobinsky models~\cite{Ellis:2013nxa, Ferrara:2013rsa} also. The ``$\alpha$-deformed'' superpotential 
\begin{align}
W=\frac{\sqrt{3\alpha}}{2} m \left( \Phi + \frac{\sqrt{3\alpha}}{2} \left( e^{- 2\Phi / \sqrt{3\alpha}} -1 \right) \right)
\end{align}
leads to the scalar potential
\begin{align}
V=\frac{3\alpha}{4}m^2 \left( 1- e^{-\sqrt{2/3\alpha}\phi} \right)^2.
\end{align}
In Appendix~\ref{sec:defStarobinsky} we demonstrate some other ways of getting the similar (or the same) deformed Starobinsky potentials generalizing those of Ref.~\cite{Ketov:2014qha}.

\paragraph{Example 3: a ``symmetry breaking'' potential.}
The ``symmetry breaking''-type potential (our inflaton is assumed to be a singlet),
\begin{align}
V=& \lambda \left( \phi^2 - v^2 \right)^2, 
\end{align}
can be used for a new~\cite{new_inflation}, chaotic~\cite{Linde:1983gd}, or topological~\cite{topological_inflation} inflation, depending on the values of the parameters and the initial conditions, see \textit{e.g.} a review~\cite{Yamaguchi:2011kg}
. It  is obtained from the following superpotential:
\begin{align}
W=& \sqrt{\lambda} \left( \frac{2}{3}\Phi^3 - v^2 \Phi  \right)~.
\end{align}

\paragraph{Example 4: a sinusoidal potential.}
The sine-modulated inflationary scalar potential $V=\frac{V_0}{2}\left( 1- \cos n \phi\right)$ for natural inflation~\cite{natural_inflation}  follows from the superpotential
\begin{align}
W= \frac{\sqrt{2V_0}}{n}\sqrt{1-\cos \sqrt{2}n \Phi } \cot \frac{n \Phi}{\sqrt{2}}~~.
\end{align}

\section{Confirmation of stabilization}\label{sec:stabilization}
To demonstrate viable stabilization, we assume a specific K\"{a}hler potential and a generic superpotential,
\begin{align}
K=& -3 \ln \left( 1+ \frac{\left(  \Phi+ \bar{\Phi} \right)+\zeta \left( \Phi +\bar{\Phi} \right)^4 }{\sqrt{3} }\right) ,\label{KA} \\
W=& \frac{1}{\sqrt{2}} \tilde{W}(-\sqrt{2}i \Phi ).
\end{align}
Note that the stabilization term (proportional to $\zeta$) does not break the shift symmetry.
Other symmetry-preserving terms, $(\Phi + \bar{\Phi})^{n}$, may appear in the K\"{a}hler potential.
In the presence of such terms, inflation can still be realized as long as the non-inflaton field is stabilized, but the resulting inflaton potential will be corrected and become complicated.
This type of the stabilization term was first introduced in Ref.~\cite{Ellis:1984bs} and recently was used in Refs.~\cite{Ellis:2013nxa, Ferrara:2014ima, Ellis:2014rxa, Ketov:2014qha}.
  Though some explanations of the mechanism are given in those references,~\footnote{
  As was pointed out to us by the referee, our setup is different from the case of two superfields, where the quartic term of the stabilizer (Polonyi) field appears in the low-energy effective theory of the O'Raifeartaigh model coupled to supergravity~\cite{Kitano:2006wz}.  In Appendix~\ref{sec:origin} we
briefly discuss the possible origin of the shift-symmetric quartic term of $\Phi$ used in our approach.
} we analyze it here again, in our specific setup, for the sake of self-completeness and transparency.
The K\"{a}hler metric is 
\begin{align}
K_{\Phi \bar{\Phi}}= \frac{1-12\sqrt{3}\zeta \left( \Phi +\bar{\Phi}\right)^2 -4\zeta \left( \Phi +\bar{\Phi}\right)^3 +4\zeta^2 \left( \Phi +\bar{\Phi}\right)^6}{\left[1+\frac{\Phi+\bar{\Phi}+\zeta \left( \Phi+\bar{\Phi} \right)^4}{\sqrt{3}}  \right]^2}~~.
\end{align}
We assume $\zeta \gtrsim \mc{O}(1)$ so that the real part (non-inflaton) must be smaller than one, in order to keep the canonical sign of the kinetic term. The scalar potential is
\begin{align}
V= & A^{-1} \left[ B^{-1} \left| W_{\Phi} \right|^2 - \sqrt{3}\left(1+8\sqrt{2}\zeta \phi^3 \right) B^{-2} \left( W_{\Phi}\bar{W}+\bar{W}_{\bar{\Phi}}W \right) \right. \cr
 & \left.+72\zeta \phi^2 \left( \sqrt{3}+\sqrt{2}\phi+4\zeta \phi^4 \right) B^{-3} \left| W \right|^2 \right]~,
\end{align}
where $\Phi=\frac{1}{\sqrt{2}}\left( \phi + i \chi \right)$ with $\phi$ and $\chi$ real, and 
\begin{align}
A=& 1-24 \sqrt{3}\zeta \phi^2 -8\sqrt{2}\zeta\phi^3+32\zeta^2\phi^6~~, \\
B=&1+\frac{\sqrt{2}\phi+4\zeta \phi^4}{\sqrt{3}}~~.
\end{align}

The expectation value of $\phi$ is obtained by truncating the higher order terms beyond  
$\mc{O}(\phi )$ in the stationary condition $V_\phi=0$,
\begin{align}
\phi=&  \frac{4\sqrt{6}\tilde{W}'(\chi)^2 -3\sqrt{6}\tilde{W}(\chi)\tilde{W}''(\chi)}{2\left( 108\sqrt{3}\zeta \tilde{W}(\chi)^2 +\left( 72\sqrt{3}\zeta+14 \right) \tilde{W}'(\chi)^2 -12\tilde{W}(\chi)\tilde{W}''(\chi) +3\tilde{W}''(\chi)^2 -3\tilde{W}'(\chi)\tilde{W}'''(\chi) \right)}  \nonumber \\
\simeq &\frac{4\sqrt{6} -3 \sqrt{3 \epsilon E}}{4\left( 54\sqrt{3}\zeta E +36\sqrt{3}\zeta -3 \sqrt{2 \epsilon E} +7 \right)} \label{phi_vev1}~~,
\end{align}
where $\epsilon= \frac{1}{2}(V'(\chi)/V(\chi))^2$ is the slow-roll parameter.
The first equality holds both during inflation and at the vacuum.
In the second equality, we have used $V\simeq |\tilde{W}'(\chi)|^2$ and have neglected the terms proportional to the slow-roll parameters, unless they are accompanied by the enhancement factor
\begin{align}
E \equiv \left( \frac{\tilde{W}(\chi)}{\tilde{W}'(\chi)} \right)^2,
\end{align}
so that it is valid during inflation.\footnote{
For simplicity of notation, we represent $\text{sgn}\left(\tilde{W}(\chi)/\tilde{W}'(\chi)\right) \sqrt{E}$ as $\sqrt{E}$ where sgn denotes the sign of the argument.
}
For example, in the monomial superpotential case, $E=\left( \frac{\chi}{n} \right)^2$ and it is large ($E>1$) during the large field inflation $(|\chi|>1)$ for $n$ of the order one.  In the case of the Starobinsky potential \eqref{W_Starobinsky}, we find $E=\left( \frac{\chi}{1-e^{-\sqrt{2/3}\chi}}-\frac{\sqrt{3}}{2} \right)^2 $, and it is also large during inflation.  Thus, typically, $E$ is of the order $\chi^2$, and $\sqrt{\epsilon E}$ is roughly of the order one.
In summary, we have
\begin{align}
\phi \simeq \mc{O} \left( 10^{-2} \zeta^{-1} E^{-1} \right) \sim \mc{O}\left (  10^{-2} \zeta^{-1} \chi^{-2} \right), \label{phi_vev2}
\end{align}
during inflation.
It is smaller than one indeed, being also consistent with the truncation above.
The kinetic term is approximately canonical.

The mass squared of $\phi$ is
\begin{align}
V_{\phi \phi}=&\frac{1}{3}  \left( 216\sqrt{3}\zeta E+ 144\sqrt{3}\zeta+ 28-12 \sqrt{2 \epsilon E} +6\epsilon -3 \eta +\mc{O}(\phi \zeta E) \right)  \tilde{W}'(\chi)^2 \nonumber \\
\simeq & 2 \zeta \left( 108\sqrt{3} E+ 72\sqrt{3} \right) H^2 ,
\end{align}
where $\eta=V''(\chi)/V(\chi)$ is another slow-roll parameter. 
The $\mc{O}\left(\phi\zeta E\right)$ term is of the order one, and  is neglected in the last expression together with other subdominant terms.
The mass of $\phi$ can be easily larger than the Hubble scale $H$. In this way the real part (non-inflaton) $\phi$ can be stabilized.

The smallness itself of $\phi$ compared to one does not ensure validity of approximation in the previous Section because the small nonzero value may break cancellation among terms in the scalar potential due to the no-scale structure.
Now we check that the corrections to the scalar potential~\eqref{potential} induced by the nonzero value of $\phi$ in eq.~\eqref{phi_vev1} are actually smaller than the leading terms.
The scalar potential (up to the leading corrections) is given by
\begin{align}
V = & \tilde{W}' \left( \chi \right) ^2 - \frac{\left( 4 \tilde{W}'\left(\chi \right)^2 - 3 \tilde{W}\left(\chi\right) \tilde{W}''\left(\chi\right)\ \right)^2}{216\sqrt{3}\zeta \tilde{W}\left(\chi\right)^2 +  \left( 144\sqrt{3}\zeta+28\right)\tilde{W}'\left(\chi\right)^2 -24\tilde{W}\left(\chi\right)\tilde{W}''\left(\chi\right)+6\tilde{W}''\left(\chi\right)^2 -6 \tilde{W}'\left(\chi\right)\tilde{W}'''\left(\chi\right)} \nonumber \\
\simeq & \tilde{W}'\left(\chi\right)^2 \left( 1  - \frac{\left( 8-3\sqrt{2\epsilon E} \right)^2}{16 \left( 54\sqrt{3}E\zeta+36\sqrt{3}\zeta-3\sqrt{2\epsilon E}+7 \right)} \right)~~. \label{VwCorrection}
\end{align}
The first equality holds both during inflation and at the vacuum, whereas the second equality is based on the same approximation in the second equality of eq.~\eqref{phi_vev1} (valid during large field inflation).
 The corrections are indeed subdominant and vanish in the limit of large $E$ (large $\chi$) or large $\zeta$ with fixed $\epsilon E$.
 For example, the numerical value of the second term in the parenthesis of eq.~\eqref{VwCorrection} is  $-2.93\times 10^{-3}$ for ($E=5$, $\epsilon=0.1$, and $\zeta=1$), and $-8.61\times 10^{-3}$ for ($E=10$, $\epsilon=0.1$, and $\zeta=0.1$).
These arguments justify our treatment of the theory as the effective single field inflationary theory where the kinetic term is approximately canonically normalized and the scalar potential is given by $V\simeq |\tilde{W}'(\chi)|^2$ in the large field regime of the inflaton $\chi$.

To convince a reader even more, we calculate numerically the dynamics of inflation.
We take two benchmark models as the examples: (i) the chaotic inflation with a quadratic potential, and (ii) the Starobinsky potential.
We set the inflaton mass and the stabilization parameter as $m=10^{-5}$ and $\zeta=1$ for simplicity.
Note that the stabilization parameter of the order one works pretty well, as is shown below.

\begin{figure}
  \begin{center}
    \includegraphics[clip, width=10cm]{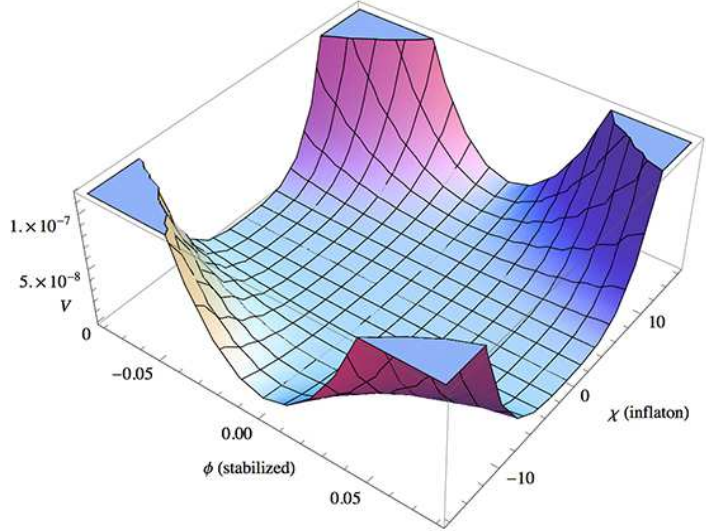}
    \caption{The scalar potential of the stabilised quadratic model. The mass scale and the stabilization strength are set to $m=10^{-5}$ and $\zeta=1$, respectively.}
    \label{fig:quadratic_potential}
  \end{center}
\end{figure}

\begin{figure}
  \begin{center}
    \includegraphics[clip, width=8cm]{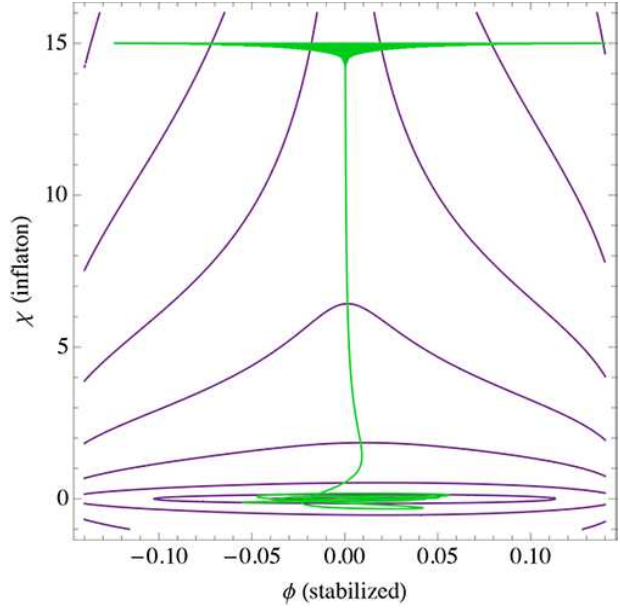}
    \caption{The inflaton trajectory (green) in the stabilized quadratic model. The initial conditions are $\phi=0.14$, $\chi=15$, $\dot{\phi}=0$, and $\dot{\chi}=0$.  The mass scale and the stabilization strength are set to $m=10^{-5}$ and $\zeta=1$,
    respectively. The contour plot of logarithm of the potential is shown in purple.}
    \label{fig:quadratic_trajectory}
  \end{center}
\end{figure}

Let us consider the example (i): the chaotic model with a quadratic potential.
The potential of the model with the stabilization proposed in this Section is shown in Fig.~\ref{fig:quadratic_potential}.
The trajectory of the inflaton field in this model is shown in Fig.~\ref{fig:quadratic_trajectory}.
We apply the initial condition away from the stabilization valley in order to check how the stabilization mechanism works. 
The real part (non-inflaton) rapidly oscillates around (damped to) the instantaneous minimum, and after that the trajectory is approximately that of single-field inflation.
It slightly deviates from the imaginary axis near the end of inflation, and finally oscillates around the vacuum.
The deviation is smaller in the larger inflaton field value (Fig.~\ref{fig:quadratic_trajectory}) because of Eq.~\eqref{phi_vev2}.
The fractional difference between the inflaton scalar potential along the trajectory and the quadratic potential is only within $1.4\%$ or even smaller well before the end of inflation.

\begin{figure}
  \begin{center}
    \includegraphics[clip, width=10cm]{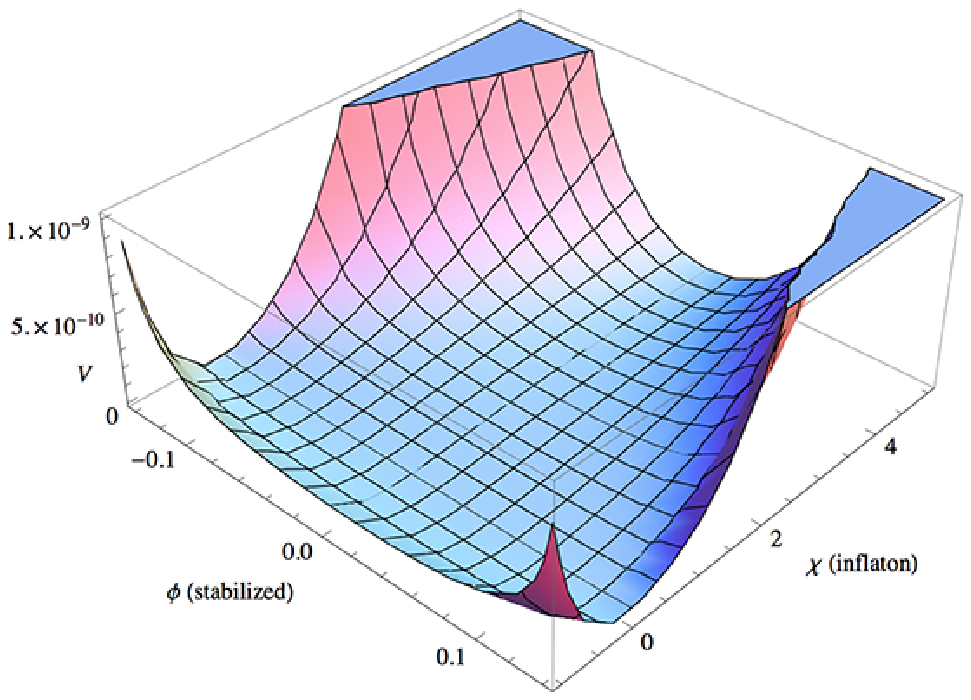}
    \caption{The scalar potential of the stabilized Starobinsky model. The mass scale and stabilization strength are set to $m=10^{-5}$ and $\zeta=1$, respectively.}
    \label{fig:Starobinsky_potential}
  \end{center}
\end{figure}

\begin{figure}
  \begin{center}
    \includegraphics[clip, width=8cm]{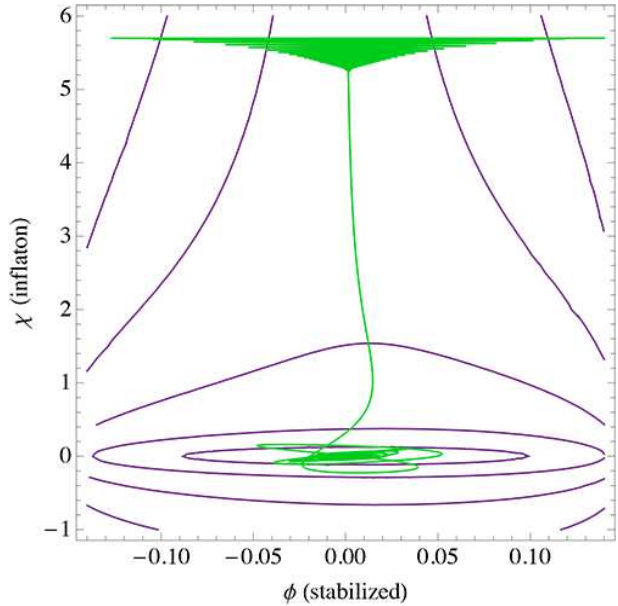}
    \caption{The inflaton trajectory (green) in the stabilized Starobinsky model. The initial conditions are $\phi=0.14$, $\chi=5.7$, $\dot{\phi}=0$, and $\dot{\chi}=0$.  The mass scale and the stabilization strength are set to $m=10^{-5}$ and $\zeta=1$, respectively. The contour plot of logarithm of the potential is shown in purple.}
    \label{fig:Starobinsky_trajectory}
  \end{center}
\end{figure}

Next, let us consider the example (ii): the Starobinsky model in our framework.
The potential of the model with the stabilization of this Section is shown in Fig.~\ref{fig:Starobinsky_potential}.
The trajectory of the inflaton field in the model is shown in Fig.~\ref{fig:Starobinsky_trajectory}.
It is qualitatively similar to the case of the quadratic model (see Fig.~\ref{fig:quadratic_trajectory}).
The fractional difference between the inflaton scalar potential along the trajectory and the Starobinsky potential is only within $2.2\%$, or even smaller well before the end of inflation.

\section{Impact of matter couplings on inflaton dynamics}\label{sec:coupling}
We realized inflation in supergravity with a single chiral superfield. After all we must couple the inflaton sector to other matter such as the Standard Model sector or a hidden sector where SUSY is broken. In this Section we consider a few simple ways of coupling and check whether they affect the inflaton dynamics.
We also discuss the inflaton decay briefly.

 We assume that the inflaton superpotential and a superpotential of other superfields are decoupled,
\begin{align}
W(\Phi, \phi^i)=W^{\text{(inf)}}(\Phi)+W^{\text{(other)}}(\phi^i),
\end{align}
where $\Phi$ is the inflaton and $\phi^i$ stand for the particle species $i$ other than inflaton.
This form is preserved during the renormalization group running due to the non-renormalization theorem~\cite{non-renormalization}.

First, let us consider the case when the K\"{a}hler potential of inflaton and that of the other fields are also decoupled,
\begin{align}
K(\Phi, \phi^i , \bar{\Phi}, \bar{\phi}^{\bar{j}})= -3 \ln \left(1+\frac{1}{\sqrt{3}}\left( \Phi +\bar{\Phi}\right) \right) +K^{\text{(other)}}(\phi^i , \bar{\phi}^{\bar{j}} ), \label{minimal_coupling}
\end{align}
where we have implicitly assumed the existence of a stabilization term under the logarithm.
This form is not preserved under the renormalization group running, but we take it  just as a simple example here.
Then the derivatives with respect to both the inflaton and the other fields vanish, so that there is no kinetic mixing between the inflaton and the other fields.
The scalar potential with the $(\Phi+\bar{\Phi})$ being stabilized at the origin is
\begin{align}
V=e^{K^{\text{(other)}}}\left( |W_{\Phi}|^2 -\sqrt{3} \left( W^{\text{(other)}}\bar{W}_{\bar{\Phi}}+\overline{W^{\text{(other)}}}W_{\Phi} \right) +K^{\text{(other)}\bar{j}i}D_i W D_{\bar{j}}\bar{W}   \right)~, \label{coupledVminimal}
\end{align}
where $D_{i}W= W_{i}+K_{i}W$ is the covariant derivative.
There are the Hubble induced masses $\sqrt{3}H$ for the scalars other than the inflaton, so light fields are frozen at the origin during inflation, whereas heavy fields are decoupled because of their own masses.
Therefore, the dynamics is essentially the single-field inflation.
If the SUSY breaking scale $W^{\text{(other)}}$ is high, the inflaton potential receives corrections.
This feature is similar to the models with several chiral superfields ---  see the discussions of SUSY breaking on inflation in these models in Refs.~\cite{Buchmuller:2014pla, Abe:2014opa}.
The impact of the conformal rescaling, required to move to the Einstein frame, on SUSY breaking term is very well controlled in the sense that the conformal factor does not depend on the inflaton because of the shift symmetry~\cite{Abe:2014opa}.
Corrections proportional to the inflaton superpotential also arise from the third term in eq.~\eqref{coupledVminimal} in the case there is a large $K_{i}$ (large VEV in the case of minimal K\"{a}hler potential).
This is in contrast to models with the sGoldstino $\vev{S}=0$ and the superpotential $W\propto S$ because the value of the superpotential vanishes in these models.
In summary, the inflaton potential is not affected by matter coupling~\eqref{minimal_coupling} if the SUSY breaking scale is low and there is no large VEV.  The latter condition is satisfied due to the Hubble-induced mass.  Further quantitative study will be done elsewhere.
Note that high-scale SUSY breaking (inflaton mass less than the mass of the SUSY breaking field) is also disfavored from the perspective of gravitino overproduction from inflaton decay --- see the text below and 
Ref.~\cite{Dine:2006ii} for more.

Inflaton can decay into gauge bosons and gauginos, if there is a coupling like
\begin{align}
\frac{1}{4}\int \text{d}^2 \Theta 2 \ms{E} \, c \Phi \mc{W}^{A}\mc{W}^{A} + \text{h.c.}~,
\end{align}
where $c$ is the coupling constant and $\mc{W}^{A}$ is the superfield strength of a real superfield.
Although this coupling breaks the shift symmetry,~\footnote{
If the coefficient $c$ is real (in our convention $\Phi$ transforms in the imaginary direction under the shift symmetry), the shift symmetry is broken only via non-perturbative effects.} it could be generated with real
$c$ as the anomaly of an underlying symmetry in the UV theory~\cite{Iwamoto:2014ywa}.
The decay rate is $N_{\text{g}} |c|^{2} m_{\chi}^{3}/128\pi M_{G}^{2}$~\cite{mod-ind gravitino}, where $N_{\text{g}}$ is the number of generators of the gauge algebra, $m_{\chi}$ is the inflaton mass, and $M_{G}$ is the reduced Planck mass.
Note that the inflaton cannot decay through the super-Weyl-K\"{a}hler and sigma-model anomaly effects~\cite{Endo:2007ih, Endo:2007sz}, because the K\"{a}hler potential does not depend on the inflaton. The
two-body decay rates into scalars and spinors are of order $m_{\chi}m^2/M_{G}^{2}$~\cite{Endo:2007sz}, where $m$ is the mass of daughter particles. 
The three-body decay rate is sizable of the order $|y_{\text{t}}|^{2}m_{\chi}^{3}/M_{G}^{2}$~\cite{Endo:2007sz}, where $y_{\text{t}}$ is the top Yukawa coupling.
The decay rate into a pair of gravitinos is $\left| \mc{G}^{\text{(eff)}}_{\chi}\right|^{2}m_{\chi}^{5}/288 \pi m_{3/2}^{2} M_{G}^{2}$~\cite{mod-ind gravitino} with the effective coupling constant in our model being evaluated as $\left| \mc{G}^{\text{(eff)}}_{\chi}\right|^{2}=27\left( \frac{m_{3/2}}{m_{\chi}} \right)^{2} \left( \frac{m_{z}^{2}}{m_{\chi}^{2}-m_{z}^{2}} \right)^{2}$~\cite{Dine:2006ii, Endo:2006tf, Endo:2012yg}, where $m_{3/2}$ is the gravitino mass, and $m_{z}$ is the mass of the SUSY breaking field $z$.

Next, let us consider the case when the K\"{a}hler potentials are summed under the logarithm,
\begin{align}
K(\Phi, \phi^i , \bar{\Phi}, \bar{\phi}^{\bar{j}})= -3 \ln \left(1+\frac{1}{\sqrt{3}}\left( \Phi +\bar{\Phi}\right) -\frac{1}{3}J(\phi^i,\bar{\phi}^{\bar{j}})\right) , \label{sequestered_coupling}
\end{align}
where $J$ is a hermitian function.
This structure can be understood \textit{e.g.}, as the geometrical sequestering of the inflaton sector and the other sectors~\cite{Randall:1998uk}.
 Again we have implicitly assumed the presence of a stabilization term. For simplicity of our notation, we introduce a function $\Omega$ such that $K=-3\ln \Omega$ or $\Omega=\exp (-K/3)$. The K\"{a}hler metric and its inverse are
\begin{align}
K_{I\bar{J}}=\Omega^{-2} \begin{pmatrix} 1 & -\frac{1}{\sqrt{3}} J_{\bar{j}} \\ -\frac{1}{\sqrt{3}}J_{i} & \Omega J_{i\bar{j}}+\frac{1}{3}J_{i}J_{\bar{j}} \end{pmatrix},  & & 
K^{\bar{J}I}=\Omega \begin{pmatrix} \Omega+\frac{1}{3}J_{i}J^i & \frac{1}{\sqrt{3}}J^{i} \\ \frac{1}{\sqrt{3}}J^{\bar{j}} & J^{\bar{j}i} \end{pmatrix},
\end{align}
where capital Latin indices $I, J,\dots$ run over $\Phi$ and $i, j, \dots$,  the $J^{\bar{j}i}$ is the inverse matrix of $J_{i\bar{j}}$, while the indices are raised and lowered by those matrices, \textit{e.g.}, $J^{i}=J^{\bar{j}i}J_{\bar{j}}$. Then the  scalar potential is given by
\begin{align}
V=&\Omega^{-2} \left( \left( \Omega +\frac{1}{3}J_{i}J^{i}\right)|W_{\Phi}|^2  -\sqrt{3} \left( W^{\text{(other)}}\bar{W}_{\bar{\Phi}} +\overline{W^{\text{(other)}}}W_{\Phi} \right)   \right. \nonumber \\
&\left.\phantom{ \left( \Omega +\frac{1}{3}J_{i}J^{i}\right) }\qquad \qquad              +\frac{1}{\sqrt{3}} \left( J^i W_i \bar{W}_{\bar{\Phi}}+J^{\bar{i}}\bar{W}_{\bar{i}}W_{\Phi} \right) +J^{\bar{j}i}W_{i}\bar{W}_{\bar{j}}  \right). \label{coupledVsequestered}
\end{align}
The Hubble induced mass is $\sqrt{2}H$, so the fields other than the inflaton are stabilized at their origin during inflation.  Assuming that these fields are charged under some unbroken symmetries, the first derivatives $J_{i}$ vanish. Then the kinetic mixing effects become negligible.  
Similar comments to the case of minimal coupling~\eqref{minimal_coupling} apply here too, but there are no terms proportional to the inflaton superpotential in Eq.~\eqref{coupledVsequestered} in the case of sequestered coupling~\eqref{sequestered_coupling} (we have again used the phase alignment condition for the inflaton superpotential).
Inflaton dynamics is not affected by matter coupling if the SUSY breaking scale is low.
 
The inflaton decay is similar to the previous example, but there are no sizable three-body decays (see Ref.~\cite{Terada:2014uia} for a similar situation).
The two-body decay rate into scalar particles is of the order $|J_{ij}|^{2}m_{\chi}^{3}/M_{G}^{2}$.
The effective coupling constant for decay into two gravitinos is $\left| \mc{G}^{\text{(eff)}}_{\chi} \right|^{2}=\left| \left( J_{z} + \frac{2 W_{z}}{m_{\chi}} \right) \frac{m_{z}^{2}}{m_{\chi}^{2}-m_{z}^{2}} \right|^{2}$.

\section{Conclusion}\label{sec:conclusion}

In this paper we proposed the simple framework for a construction of arbitrary positively 
semi-definite single field inflationary potentials in supergravity by using only a single chiral superfield.
The scalar potential --- see Eqs.~\eqref{potential} and \eqref{potential2} ---  in our framework has (approximately) the same form as the $F$-term in global SUSY theory, and it effectively becomes a function of a single field (inflaton) due to the stabilization mechanism. The inflaton does not break SUSY at the vacuum. We verified that our stabilization works, and we also proposed some simple ways of coupling the inflaton sector to other matter sectors, without affecting the inflationary dynamics.

Our class of the very economical models provides vast possibilities for realizing cosmological inflation in supergravity, which are consistent with the observational data in the most minimal setup (with a single chiral superfield).

\section*{Acknowledgements}
SVK was supported by a Grant-in-Aid of the Japanese Society for Promotion of Science (JSPS) under No.~26400252, by the World Premier International Research Centre Initiative (WPI Initiative), MEXT, Japan, and by the Competitiveness Enhancement Program of the Tomsk Polytechnic University in Russia. TT is grateful to colleagues at the University of Tokyo, especially to Koichi Hamaguchi and Kazunori Nakayama, and also to Shuntaro Aoki and Yusuke Yamada, for useful discussions. TT was supported by a Grant-in-Aid for JSPS Fellows, and a Grant-in-Aid of the JSPS under No.~2610619.

\appendix

\section{Specific stabilization for various K\"{a}hler potentials}\label{sec:VariousK}

In this paper we focused on the special K\"{a}hler potential and a class of the superpotentials such that an arbitrary scalar potential is readily available, but there are still many possibilities leading to a single field inflation from supergravity with a single chiral superfield, as was already explained in Ref.~\cite{Ketov:2014qha}. Here we examine the stabilization quality in those theories, and justify our treatment.

Let us begin with the ``minimal'' K\"{a}hler potential (see Eq.~(9) in Ref.~\cite{Ketov:2014qha}),
\begin{align}
K=\frac{1}{2} \left( \Phi + \bar{\Phi} \right)^{2} - \zeta \left( \Phi + \bar{\Phi} - 2\Phi_0 \right)^{4}.
\end{align}
The inflaton is $\text{Im}\Phi$, and $\text{Re}\Phi$ is stabilized by the $\zeta$ term.
The scalar potential for a general superpotential $W(\Phi)$ reads
\begin{align}
V=& \frac{e^{2\Phi_0^2-\zeta \left( 2 \text{Re}\Phi - 2\Phi_0 \right)^4}}{1-12\zeta \left( 2 \text{Re}\Phi - 2\Phi_0 \right)^2 } \left( |W_{\Phi}|^2 + 2\left( \text{Re}\Phi-2\zeta \left( 2 \text{Re}\Phi - 2\Phi_0 \right)^3  \right) \left( W_{\Phi}\bar{W}+\bar{W}_{\bar{\Phi}}W\right)  \right. \nonumber \\
& \qquad  \left.  +4 \left( \text{Re}\Phi-2\zeta \left( 2 \text{Re}\Phi - 2\Phi_0 \right)^3  \right)^2 |W|^2 \right) -3e^{2\Phi_0^2-\zeta \left( 2 \text{Re}\Phi - 2\Phi_0 \right)^4}|W|^2~.
\end{align}

If a deviation of $\vev{\text{Re}\Phi}$ from $\Phi_0$ is small, it merely results in a small change of the coefficient at each term in the above expression, so that the inflaton dynamics receives only a minor change. In fact, the expectation value of $\text{Re}\Phi$ around $\Phi_0$ is found to be small indeed, similarly to the analysis in Sec.~\ref{sec:stabilization},
\begin{align}
\vev{\text{Re}\Phi}-\Phi_0\simeq - \frac{\Phi_0 \left( 4\Phi_0^2-1 \right) }{96\zeta \Phi_0^2+16\Phi_0^4+8\Phi_0^2-1}\simeq \mc{O}\left(10^{-1}\zeta^{-1}\right) \label{dev_minK}
\end{align}
for $\Phi_0\simeq \mc{O}(1)$, where we have neglected the subdominant powers in $\text{Im}\Phi$ (inflaton) on dimensional reasons, $|W_{\Phi}|\sim |W /\Phi|$, $|W_{\Phi\Phi}|\sim |W/\Phi^2|$, \textit{etc}. In this case, a relatively large $\zeta$ is required to suppress the deviation~\eqref{dev_minK}, \textit{e.g.} $\zeta\simeq \mc{O}(10)$ for $\vev{\text{Re}\Phi}-\Phi_0\simeq \mc{O}(10^{-2})$. It is worth noticing that a correction to the kinetic term, $12 \zeta (2\text{Re}\Phi-2\Phi_0)^2$, is also suppressed to be $\mc{O}(10^{-1}\zeta^{-1})$. The mass squared of the non-inflaton is
\begin{align}
V_{\phi \phi}\simeq& \frac{6\left(16 \Phi_0^4+\left( 96\zeta +8 \right)\Phi_0^2-1  \right)}{4\Phi_0^2-3}H^2,
\end{align}
where $\Phi=(\phi+i \chi)/\sqrt{2}$, and  we have used the slow-roll Friedmann equation, $V\simeq 3 H^2$, and $V\simeq e^{2\Phi_0^2}(4\Phi_0^2-3)|W|^2$.
This mass can be larger than the Hubble scale with moderate values of $\Phi_0$ and $\zeta$.

Next, let us consider the logarithmic K\"{a}hler potential (see Eq.~(25) of Ref.~\cite{Ketov:2014qha}),
\begin{align}
K=-3\ln \left[ \frac{\Phi+\bar{\Phi}+\zeta \left( \Phi+\bar{\Phi}-2\Phi_{0} \right)^{4}}{3}\right]~.
\end{align}
The inflaton is $\text{Im}\Phi$, and $\text{Re}\Phi$ is stabilized by the $\zeta$ term.
This is related to Eq.~\eqref{KA} via field redefinition, but we do not make assumptions about a superpotential here. The scalar potential is (see Eq.~(26) in Ref.~\cite{Ketov:2014qha})
\begin{align} 
V & = \frac{9}{\left[\Phi+\bar{\Phi}+\zeta (\Phi+\bar{\Phi}-2\Phi_0)^4 \right]^2}\nonumber \\
& \times \frac{1}{  1-4\zeta (\Phi+\bar{\Phi}-2\Phi_0 )^{3}+4\zeta ^2 (\Phi+\bar{\Phi}-2\Phi_0 )^{6}-24\zeta \Phi_0 \left( \Phi+\bar{\Phi}-2\Phi_{0} \right)^{2} } \nonumber \\
&\times  \left[ (\Phi+\bar{\Phi}+\zeta (\Phi+\bar{\Phi}-2\Phi_0)^4 )\left|W_{\Phi}\right|^{2}- 3\left( 1+4\zeta (\Phi+\bar{\Phi})^{3}\right) \left( \bar{W}W_{\Phi}+W\bar{W}_{\bar{\Phi}}\right)  \right.\nonumber \\
& \left. \phantom{ (\Phi+\bar{\Phi}+\zeta (\Phi+\bar{\Phi}-2\Phi_0)^n \left|W_{\Phi}\right|^{2}}   +108\zeta (\Phi+\bar{\Phi}-2\Phi_0 )^{2}|W|^{2} \right]~.
\end{align}
After the canonical normalization of the real part, $\text{Re}\Phi=\Phi_0 e^{\sqrt{2/3}\phi}$, the deviation is evaluated as
\begin{align}
\vev{\phi} \simeq \frac{\sqrt{6}\left( W\bar{W}_{\bar{\Phi}}+\bar{W}W_{\Phi}  \right)}{288\Phi_0^2 \zeta |W|^2}\sim \mc{O}(10^{-2}\zeta^{-1}\text{Im}\Phi^{-1}).
\end{align}
This gives rise to the term proportional to $|W|^2$ which is absent in the ideal case $\text{Re}\Phi=\Phi_0$, and is actually subdominant here. The mass squared of the non-inflaton is
\begin{align}
V_{\phi\phi}\simeq 1296 \zeta |W|^2, 
\end{align}
so that there is no difficulty to make the mass larger than the Hubble scale.

Finally, let us consider the K\"{a}hler potential used in Sec.~3 of Ref.~\cite{Ketov:2014qha},
\begin{align}
K=-3\ln \left[ \frac{\Phi+\bar{\Phi}+\zeta \left(-i \Phi+i \bar{\Phi}-2\Phi_{0} \right)^{4}}{3}\right]~.
\end{align}
In this case the imaginary part $\text{Im}\Phi$ is stabilized and the real part $\text{Re}\Phi$ is used as the inflaton. The scalar potential is
\begin{align}
V & = \frac{9}{\left[\Phi+\bar{\Phi}+\zeta (-i\Phi+i\bar{\Phi}-2\Phi_0)^4 \right]^2}\nonumber \\
& \times \frac{1}{  1-12\zeta \left( \Phi+\bar{\Phi} \right)(-i\Phi+i\bar{\Phi}-2\Phi_0)^2 +4 \zeta^2  \left(-i \Phi+i\bar{\Phi}-2\Phi_{0} \right)^{6} } \nonumber \\
&\times  \left[ (\Phi+\bar{\Phi}+\zeta (-i\Phi+i\bar{\Phi}-2\Phi_0)^4 ) \left|W_{\Phi}\right|^{2}- 3 \left( \bar{W}W_{\Phi}+W\bar{W}_{\bar{\Phi}}\right)   \right.\nonumber \\
& \left. +12i\zeta (-i\Phi+i\bar{\Phi}-2\Phi_0)^3 \left( W\bar{W}_{\bar{\Phi}}-\bar{W}W_{\Phi} \right)  +108\zeta (-i\Phi+i\bar{\Phi}-2\Phi_0 )^{2}|W|^{2} \right]~.
\end{align}
The deviation of $\text{Im}\Phi$ is obtained as
\begin{align}
\vev{\text{Im}\Phi} -\Phi_0 \simeq & i \frac{-3 \left( \bar{W}W_{\Phi\Phi}-W\bar{W}_{\bar{\Phi}\bar{\Phi}} \right) +2\text{Re}\Phi \left( \bar{W}_{\bar{\Phi}}W_{\Phi\Phi}-W_{\Phi}\bar{W}_{\bar{\Phi}\bar{\Phi}} \right) }{864\zeta |W|^2 -576\zeta \text{Re}\Phi \left( \bar{W}W_{\Phi}+W\bar{W}_{\bar{\Phi}} \right) +384 (\text{Re}\Phi)^2\zeta |W_{\Phi}|^2 } \nonumber \\
\sim& i \mc{O}(10^{-2}\zeta^{-1}(\text{Re}\Phi)^{-2}).
\end{align}
This ensures that the correction terms are subdominant again. The mass squared of the imaginary part $\chi=\sqrt{2/3}2\Phi_0^2\text{Im}\Phi$ is
\begin{align}
V_{\chi\chi}\simeq 3\zeta \left( 864|W|^2-576\text{Re}\Phi \left( \bar{W}W_{\Phi}+W\bar{W}_{\bar{\Phi}} \right)+384 \left(\text{Re}\Phi\right)^2|W_{\Phi}|^2  \right), 
\end{align}
and appears to be  $\mc{O}(10^2 \zeta (\text{Re}\Phi)^3)$ times larger than the Hubble scale squared.

According to this Appendix and Sec.~\ref{sec:stabilization}, the non-inflaton field can be strongly stabilized with the parameter $\zeta$ that is not much larger than one, which justifies our basic demand for the inflationary supergravity model to be treated as a single-field inflation.

\section{Supergravity realizations of the deformed Starobinsky models}\label{sec:defStarobinsky}

In this Appendix we demonstrate the other two ways (i.e. different from that in Sec.~\ref{sec:ArbPot}) to obtain the deformed Starobinsky potentials generalizing the models described in Ref.~\cite{Ketov:2014qha}. 

\begin{figure}[htbp]
  \begin{center}
    \includegraphics[clip,width=8cm]{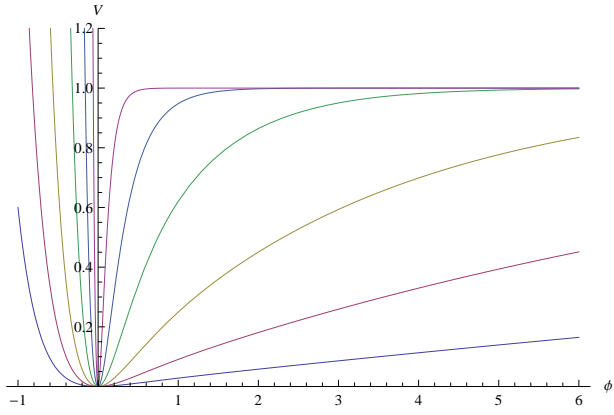}
    \caption{The deformed Starobinsky potential~\eqref{Vnp} derived from the superpotential containing a negative power~\eqref{Wnp}. The powers are $-0.03, -0.1, -0.3, -1, -3$, and $-10$ from the bottom to the top. The parameter values are taken as $a=-b=-c=d=1$.}
    \label{fig:NegativePower}
  \end{center}
\end{figure}

First, let us recall Eq.~(32) of that paper, where we employed the no-scale K\"{a}hler potential and the superpotential containing a term with a negative power $-n=-1$. Let us now generalize it to an arbitrary negative power as follows:
\begin{align}
K=& -3 \ln \left[ \left(\Phi +\bar{\Phi} \right)/3 \right] , \\
W=& \frac{1}{n}c_{-n}\Phi^{-n}+c_0+\frac{1}{3}c_3 \Phi^3~. \label{Wnp}
\end{align}
After stabilization of the imaginary part, the scalar potential becomes
\begin{align}
V= a + b e^{-n \phi}+ c e^{-(n+3)\phi} + d e^{-(2n+3)\phi}~, \label{Vnp}
\end{align}
where $\phi=\sqrt{3/2} \ln \text{Re}\Phi$ is the canonically normalized inflaton, and $a=-27\text{Re}(c_0 \overline{c_3})/2$, $b=-9(1+3/n) \text{Re}(c_{-n}\overline{c_3}) /2$, $c=27 \text{Re}(c_0 \overline{c_{-n}})/2$, and $d=9(1+3/n)|c_{-n}|^2/2$.  The potential is shown in Fig.~\ref{fig:NegativePower}.
It is possible to choose $a=d=-b=-c>0$, which ensures $V=0$ in the vacuum $\phi=0$.

\begin{figure}[htbp]
  \begin{center}
    \includegraphics[clip,width=8cm]{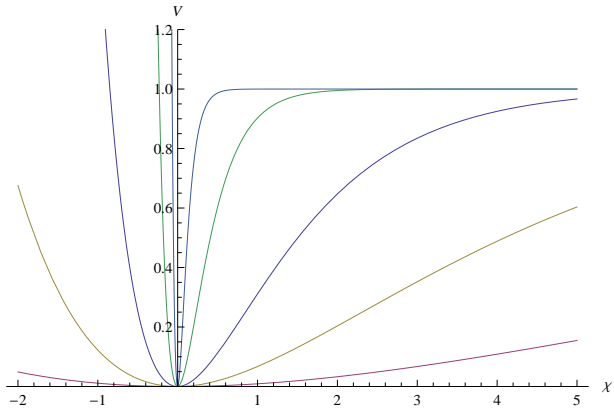}
    \caption{The deformed Starobinsky potential~\eqref{Vdeformed} derived from the superpotential~\eqref{Wdeformed}. The parameter $a$ is set to $-0.1, -0.3, -\sqrt{2/3}$ (Starobinsky), $ -3$, and $-10$ from bottom to top. The height of the potential is normalised to one.}
    \label{fig:deformed}
  \end{center}
\end{figure}

Finally, let us vary the parameter $a$ of Eq.~(34) in Ref.~\cite{Ketov:2014qha}.
The K\"{a}hler potential is taken to be the minimal one with the shift symmetry. We redefine the normalization of $a$ here and take
\begin{align}
W=m\left[ b-e^{i\sqrt{2}a (\Phi -\Phi_0 ) } \right]. \label{Wdeformed}
\end{align}
The scalar potential for $\chi= \sqrt{2} \text{Im}\Phi$ is then given by
\begin{align}
|m|^{-2}e^{-2\Phi_0^2}V=\left( 4\Phi_0^2-3\right)\left( \text{Re}b-e^{-a\chi} \right)^2 +\left( 2\Phi_0 \text{Im}b -\sqrt{2}a e^{-a\chi}\right)^2 -3(\text{Im}b)^2.
\end{align}
There exist solutions for $\text{Re}b$ and $\text{Im}b$ that lead to the scalar potential
\begin{align}
V=e^{2\Phi_0^2}|m|^2 \left( 4\Phi_0^2-3+2a^2 \right) \left( 1- e^{-a \chi } \right)^2~. \label{Vdeformed}
\end{align}
The shape of this potential is displayed in Fig.~\ref{fig:deformed}.

\section{On the possible origin of the quartic term}\label{sec:origin}

The quartic term in the K\"{a}hler potential plays the key role in stabilization of the non-inflaton scalar of the inflaton superfield in our class of models.  The simplest interpretation of the quartic term may be by assuming 
its presence at the tree level. The quartic term respects the shift symmetry, while all kinds of terms allowed by the symmetries of the theory should generically be included into the effective field theory. The remaining questions are (i) how much the quadratic and cubic terms have to be suppressed,  and (ii) what is the mechanism for their suppression. It requires a separate investigation. However, as a preliminary test, we find that small quadratic and/or cubic terms destroy the cancellation of the no-scale type model.
Hence, our K\"{a}hler potential should be regarded as a tuned one, in order to suppress the quadratic and cubic terms.  Without such tuning, the theory describes more general inflationary models in supergravity~\cite{Ketov:2014qha}.

The quartic term may also originate from some UV-completion of our phenomenological supergravity description, such as superstring theory.  Unfortunately, exact superstring corrections are not available in the
literature. Though a thorough study of the origin of the quartic term from superstring theory is beyond the scope of this paper, we briefly discuss the possible origin of the quartic term in the QFT framework, which serves as an existence proof.

In the following, we consider coupling of the inflaton to other superfields, and discuss the effective terms to be obtained through integrating out these superfields. The idea is to give the $(\Phi+\bar{\Phi})$-dependent masses to the other superfields.
The quantum correction to the frame function in Jordan frame is known, whose expression depends on the masses of the fields in the Jordan frame.
After integrating out these fields, the quantum corrected frame function $\Omega=-3\exp \left(-K\left(\Phi+\bar{\Phi}\right)/3 \right)$ is left.
See Appendix B of Ref.~\cite{Lee:2010hj} for the case of the quartic stabilization term for the stabilizer field.

The quartic term must preserve shift symmetry, so we cannot introduce a coupling to the superpotential that is holomorphic.~\footnote{An exception could be the form of $W\sim e^{\Phi}W_0 (X)$ (shift of $\Phi$ changes the phase of the superpotential; $X$ collectively denotes other superfields). However, $\Phi$ can be moved into the 
K\"{a}hler potential by a K\"{a}hler transformation, $K\sim K_{\text{original}}+\Phi+\bar{\Phi}$ and $W\sim W_0 (X)$.}
Therefore, we consider a coupling in the K\"{a}hler potential.
Let us suppose the following K\"{a}hler potential:
\begin{align}
K=-3\ln \left( -\Omega /3 \right) = -3 \ln \left( 1+\frac{1}{\sqrt{3}}\left(\Phi+\bar{\Phi}\right) -\frac{1}{3} A(\Phi+\bar{\Phi})J (X,\bar{X})    \right), \label{non_min_coupling}
\end{align}
where $A=A(\Phi+\bar{\Phi})$ is a function of $\Phi+\bar{\Phi}$, and $J$ is the kinetic function for other superfields.
If $\Phi$ breaks SUSY, it gives $X$ mass by the term like $A'' |F^{\Phi}|^2 |X|^2$, but we do not want SUSY to be broken above the inflation scale.

Let us take a simple superpotential for $X$, $W=\frac{1}{2}m X^2$, where $m$ is a mass parameter much larger than the inflaton mass.
Masses squared divided by the coefficient of the kinetic terms for scalar and spinor particles are approximately 
$m_{0}^2=\Omega^2 m^2/ 9A$ and $m_{1/2}^2=\Omega^3 m^2 / 27 A^2$, respectively, in the Jordan frame.
Then the one-loop correction to the frame function $\Omega$ is \cite{Lee:2010hj} 
\begin{align}
\Delta \Omega = -\frac{1}{16\pi^2} \left( \left( 1- A \right) m_{0}^2 \ln \left( \frac{m_{0}^2}{\mu^2} \right)   + m_{1/2}^2 \ln \left( \frac{m_{1/2}^2}{\mu^2}  \right) \right).
\end{align}

After expanding $A$ as $A\left(\Phi+\bar{\Phi}\right)=1+c_{1}\left(\Phi+\bar{\Phi}\right)+c_{2}\left(\Phi+\bar{\Phi}\right)^2+\dots$, one can easily see that the quartic term appears, as well as the higher and lower order terms.
The full expression is long and is not illuminating. The higher order terms do not affect our inflationary dynamics as far as it is expanded around the VEV. The zeroth and first order terms correspond to renormalization of the Newton constant and the field. The coefficients of quadratic and cubic terms can be eliminated at some renormalization scale $\mu$ by tuning $c_{2}$ and $c_{3}$ in the non-minimal coupling function $A\left(\Phi+\bar{\Phi}\right)$.

Hence, as was anticipated above, it is possible to obtain the quartic term in $\left(\Phi+\bar{\Phi}\right)$ from quantum corrections of heavy fields, with some tuning to suppress the unwanted terms. The origin of the non-minimal coupling in Eq.~\eqref{non_min_coupling} should be sought in an UV-complete framework.



\end{document}